\documentclass{article}
\usepackage{arxiv}
\usepackage{graphicx} 
\usepackage{multicol}
\usepackage{float}
\usepackage{mwe}
\usepackage{xcolor}
\usepackage{appendix}
\usepackage{amsmath}
\usepackage{amssymb}
\usepackage{hyperref}       
\usepackage{authblk}
\usepackage[
backend=biber,
style=alphabetic,
]{biblatex}
\usepackage{chngcntr} 
\usepackage[normalem]{ulem}

\definecolor{linkcol}{HTML}{024d6e}
\definecolor{correction_color}{HTML}{ff306c}
\hypersetup{
    colorlinks=True,
    linkcolor=linkcol,
    citecolor=linkcol,
    urlcolor=linkcol}

\newcommand{\avg}[1]{\langle #1 \rangle}
\newcommand{\diff}[1]{\mathrm{d}#1\!\mathop{}}
\newcommand{\mcom}{\ ,}
\newcommand{\mdot}{\ .}
\renewcommand{\phi}{\varphi}
\newcommand{\mbf}[1]{\mathbf{#1}}
\newcommand{\du}[1]{\underline{\underline{#1}}}
\newcommand{\overbar}[1]{\mkern 1.5mu\overline{\mkern-1.5mu#1\mkern-1.5mu}\mkern 1.5mu}

\title{Effective Energy, Interactions and Out of Equilibrium nature of Scalar Active Matter}

\author[1,$\dagger$]{Antonin Brossollet}
\author[2,$\dagger$]{Etienne Lempereur}
\author[3,4]{Stéphane Mallat}
\author[1]{Giulio Biroli}
\affil[1]{Laboratoire de Physique de l'\'Ecole normale sup\'erieure, ENS, Universit\'e PSL, CNRS, Sorbonne Universit\'e, Universit\'e de Paris F-75005 Paris, France}
\affil[2]{Département d’Informatique, ENS, Université PSL, Paris, France}
\affil[3]{Collège de France, Paris, France}
\affil[4]{Flatiron Institute, New York, USA}
\affil[$\dagger$]{\textit{These authors contributed equally to this work}}

\begin{document}
\maketitle
\begin{abstract}
Estimating the effective energy, $E_\text{eff}$ of a stationary probability distribution is a challenge for non-equilibrium steady states. Its solution could offer a novel framework for describing and analyzing non-equilibrium systems. In this work, we address this issue within the context of scalar active matter, focusing on the continuum field theory of Active Model B+. We show that the Wavelet Conditional Renormalization Group method allows us to estimate the effective energy of active model B+ from samples obtained by numerical simulations. We investigate the qualitative changes of $E_\text{eff}$ as the activity level increases. Our key finding is that in the regimes corresponding to low activity and to standard phase separation the interactions in $E_\text{eff}$ are short-ranged, whereas for strong activity the interactions become long-ranged and lead to micro-phase separation. By analyzing the violation of Fluctuation-Dissipation theorem and entropy production patterns, which are directly accessible within the WCRG framework, we connect the emergence of these long-range interactions to the non-equilibrium nature of the steady state.  This connection highlights the interplay between activity, range of the interactions and the fundamental properties of non-equilibrium systems.    
\end{abstract}
\begin{multicols}{2}

\section{Introduction}
Equilibrium statistical physics is a well-understood field with powerful tools available to thoroughly describe any system in equilibrium \cite{chaikin_principles_1995}. Fundamentally, equilibrium is associated with the time reversal symmetry of the dynamics or in the language of Markov chains, with detailed balance (or micro-reversibility) \cite{tauber_critical_2014}.  One of the most important consequences of this property is the strong link between dynamics and statics. In fact, the forces driving the dynamical behavior derive from an energy function $E$. The steady state reached at long times is characterized by the Boltzmann-Gibbs distribution with the same energy function $E$.

However, many, if not most, real-world systems operate away from equilibrium: detailed balance is violated, and the link between dynamics and statics does not hold anymore. While small deviations from equilibrium can be effectively described using for instance linear response theory and perturbation theory \cite{martin_statistical_2021}, systems far from equilibrium pose significant challenges because traditional tools are not easily applicable. The main difficulty is that even though forces and dynamical equations are known,  there are no general principles allowing to determine the steady state probability distribution.  One can define an effective energy function, $E_\text{eff}$, as minus the logarithm of the steady state distribution,  but except for very few cases, e.g. \cite{derrida_large_2002}, this effective energy  cannot be obtained analytically and its general properties are unknown.  Contrary to equilibrium, there is no more a simple relationship between the forces driving the dynamics  and $E_\text{eff}$.  For instance,  even if the forces are short-ranged, there is no guarantee for $E_\text{eff}$ to be short-ranged. Indeed, $E_\text{eff}$ has been shown to contain medium and long-range interactions in the few cases in which an exact analysis was achieved \cite{derrida_large_2002}. These interactions could play a major role in giving rise to the new and striking phenomena observed in out of equilibrium systems such as, e.g., flocking \cite{cavagna_physics_2018}, motility induced macrophase \cite{cates_motility-induced_2015} and microphase separation \cite{caporusso_motility-induced_2020}, and turbulence \cite{wallace_space-time_2014}.
In consequence,  characterizing the energy function,  and its interactions across scales, is a major theoretical challenge.

This paper aims to address this challenge by employing the Wavelet Conditional Renormalization Group (WCRG) method \cite{marchand_multiscale_2023,guth_conditionally_2023,lempereur_hierarchic_2024}, to construct $E_\text{eff}$ from realizations of the system. WCRG uses wavelet theory to decompose samples at different scales, allowing for a well-conditioned inference of the parameters of a parametric energy model.  We apply this method to obtain the effective energy description of Active Model B+ (AMB+), which is a general continuum field theory describing isotropic bulk phases of self-propelled particles \cite{wittkowski_scalar_2014}, such as microorganisms or synthetic micro-swimmers. This model is a generalization of model B, which was introduced to analyze the dynamics of equilibrium critical phenomena \cite{hohenberg_theory_1977}. It offers a suitable framework for our work, as its study is interesting on its own, and at the same time it offers a challenging but simple and general enough setting for the application of WCRG to non-equilibrium systems. 
The outcome of the WCRG analysis is a quantitatively accurate model of the effective energy, which identifies interactions across scales in terms of multi-scale long-range scalar potentials. Our results establish a link between emergent medium-long range interactions in $E_\text{eff}$ and the non-equilibrium nature of the steady states. In fact, we show that these interactions are responsible for the micro-phase separation induced by activity, and that their range is directly connected to the scale associated to entropy production patterns and fluctuation-dissipation theorem violations.

The emergence of long-range correlations, and possibly long-range interactions, is a central feature of non-equilibrium systems, studied in \cite{tailleur_statistical_2008,baek_generic_2018,arnoulx_de_pirey_anomalous_2024,ben_dor_disordered_2022,fily_athermal_2012,granek_colloquium_2024,granek_bodies_2020,ni_tunable_2015,redner_structure_2013,ro_disorder-induced_2021,cates_motility-induced_2015} for active systems. It is also a major difficulty for inferring effective energies, as it leads to singular estimation problems \cite{marchand_multiscale_2023}. Our approach leverages wavelet decomposition and multi-scale decomposition of the probability law to stabilize the estimation problem, opening the way to a characterization of the long-range interactions leading to large scale phenomena associated to non-equilibrium systems.

\section{Wavelet-Conditional RG and Effective Energy Models}
\label{subsec:WCRG}
As discussed in the introduction, here we are interested in using data to estimate the steady state distribution, and the effective energy, of many-body out of equilibrium systems. One of our aims is to show how this task can be achieved by the Wavelet Conditional Renormalization Group (WCRG) \cite{marchand_multiscale_2023,guth_conditionally_2023,lempereur_hierarchic_2024}. WCRG is a data-driven method which proceeds as an inverse renormalization group procedure: it goes from large scales down to microscopic scales, and it gains information from data at each scale to eventually reconstruct the microscopic effective energy.

\begin{figure}[H]
    \centering
    \includegraphics[width=.95\linewidth]{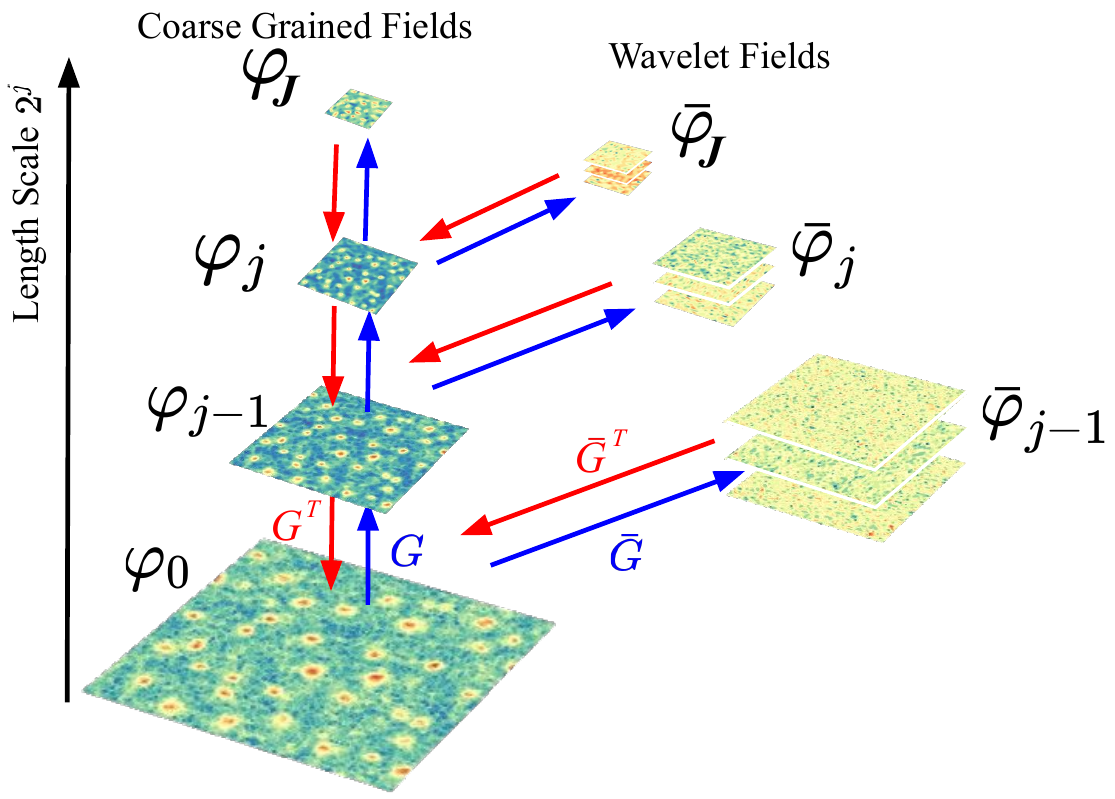}
    \caption{ The iterative wavelet decomposition, or fast wavelet transform, (in blue) iteratively decomposes the field $\varphi_{j-1}$,with length scale $2^{j-1}$, into a coarser approximation $\varphi_j$, and 3 wavelet coefficient images $\overbar{\varphi}_j$, with sub-sampled low-pass filtering $G$ and high-pass filtering, along different directions, $\overbar{G}$. It can be inverted (in red), and $\varphi_{j-1}$ can be recovered from $(\varphi_j,\varphi_{j-1})$, with the transpose operators $(G^{\rm T}\overbar{G}^{\rm T})$. Illustration is done over a realisation of AMB+ at parameters $ (\lambda, \xi) = (1,4)$. }
    \label{fig:WCRG}
\end{figure}

The procedure is illustrated in Fig. \ref{fig:WCRG} which shows the iterative wavelet decomposition of a initial physical field $\phi_0$ into a coarse grained field $\phi_J$ and wavelets fields $\overbar{\phi}_j$. This decomposition allows to rewrite the full probability law, $p_0(\phi_0)$, in a multi-scale way: 
\begin{equation}
    \label{eq:condiproba}
    p_0(\phi_0) = p_J(\phi_J) \prod_{j=1}^{J} \overbar{p}_j(\overbar{\phi}_j | \phi_j) \mdot
\end{equation}
WCRG focuses directly on the conditional probability distribution for the wavelet fields $\overbar{p}_j(\overbar{\phi}_j | \phi_j)$ at each scale. Their estimation is numerically stable and their sampling is fast because at each scale they only contain high frequencies \cite{marchand_multiscale_2023}. WCRG can then be used for systems exhibiting long-range correlations, for which a direct estimation of the effective microscopic energy would fail. The model for the conditional probabilities is low dimensional, in some cases nearly Gaussian \cite{guth_conditionally_2023,lempereur_hierarchic_2024}, and is physics-informed.

Once all the conditional probability are estimated, one has a full knowledge of the probability distribution $p_0(\phi_0)$. One can then sample new data efficiently, as shown in \cite{marchand_multiscale_2023}. Even more importantly one has access to the interactions between degrees of freedom across scales. In particular, using the model for $\overbar{p}_j(\overbar{\phi}_j | \phi_j)$ introduced in \cite{marchand_multiscale_2023}, one can obtain the effective energy for the microscopic field $\phi_0$ as
\begin{equation}
\label{eq:ansatz}
    E_\text{eff}(\varphi) = \varphi_0^{\rm T}K_0 \varphi_0 +\sum_{j=0}^J V_j(\varphi_j) \mdot
\end{equation}
The first term is a quadratic non-local interaction. In simple cases it reduces to the discrete version of a short-range differential operator, e.g. a Laplacian. The $V_0$ term is a local potential contribution $V_0(\varphi_0)=\sum_x V_0(\varphi_0(x))$. The other terms are analogous to $V_0$ but act on the coarse-grained versions of the field (and the sum over $x$ runs on the coarse-grained lattice). They represent progressively longer-range interactions, which as we shall show play an important role out of equilibrium. By construction, they do not contain any linear and quadratic term to avoid redundancy with the two previous terms.

The estimation of the conditional probabilities on each scale can be performed in parallel. Rather than minimizing the Kullback-Leibler (KL) divergence, which is computationally expensive, we minimize the score (the gradient of the logarithm of the probability distribution), which is much easier to do. Technical details are laid out in the Methods section and in depth in Appendix C of \cite{lempereur_hierarchic_2024}. A noteworthy point is that WCRG can be adapted to the regularity of the field by choosing the suitable type of wavelet, as discussed in more details in the Methods section.

\section{Active Model B+}\label{subsec:AMB}
Active matter has been one of the most studied research topic in non-equilibrium physics in recent years \cite{vrugt_review_2024,gompper_2020_2020,das_introduction_2020}. Active matter systems are studied in biology, soft-matter physics and animal behaviors. They are formed by several interacting microscopic degrees of freedom (bacteria, cells, vibrated particles, insects, birds,..,) which use and dissipates energy to move or exert mechanical forces. In consequence, active systems violates microscopic reversibility (local detailed balance) and are intrinsically out of thermal equilibrium. They display a wide diversity of new behaviors compared to their equilibrium counterparts. In order to study the change due to activity on critical and collective behaviors, several works have studied active versions of equilibrium models introduced for critical dynamics \cite{hohenberg_theory_1977, tauber_critical_2014}. In this work, we focus on one of the most studied systems: Active Model B+ (AMB+) \cite{tjhung_cluster_2018}. As the name suggests it is an generalization of the well known Model B \cite{hohenberg_theory_1977, tauber_critical_2014}, describing the stochastic equilibrium dynamics of the conserved scalar $\phi^4$ field theory. AMB+ is obtained by adding to it all terms allowed by symmetries up to order $\mathcal{O}(\nabla^4 \phi^2)$
\begin{equation}
\begin{gathered}\label{eq:amb_evo}
    \partial_{t} \phi = - \nabla \cdot (\mathbf{J}  + \sqrt{2DM} \mathbf{\Lambda} ) \\
    \mathbf{J}/M = - \nabla \left[ \frac{\delta \mathcal{F}}{\delta \phi} + \lambda |\nabla \phi|^2  \right] + \xi ( \nabla^2 \phi) \nabla \phi \\
    \mathcal{F}[\phi] = \int{ \left\{ -\frac{\phi^2}{2} + \frac{\phi^4}{4} + \frac{K}{2} |\nabla \phi|^2 \right\} } \diff{\mathbf{x}} \\
    \avg{ \Lambda^{m}(\mathbf{x},t) \Lambda^n(\mathbf{x'},t')} = \delta^{mn} \delta(\mathbf{x}-\mathbf{x'}) \delta(t-t') \mcom
\end{gathered}
\end{equation}
where $\Lambda(\mathbf{x},t)$ is Gaussian white noise with zero mean and unit variance, and $D$ is the strength of thermal fluctuations (proportional to the temperature). This is the most general model for a conserved field at this order. In the following, for simplicity, we set the coefficients $K$ and $M$ to constants equal to one.

\begin{figure}[H]
    \centering
    \includegraphics[width=1.\linewidth]{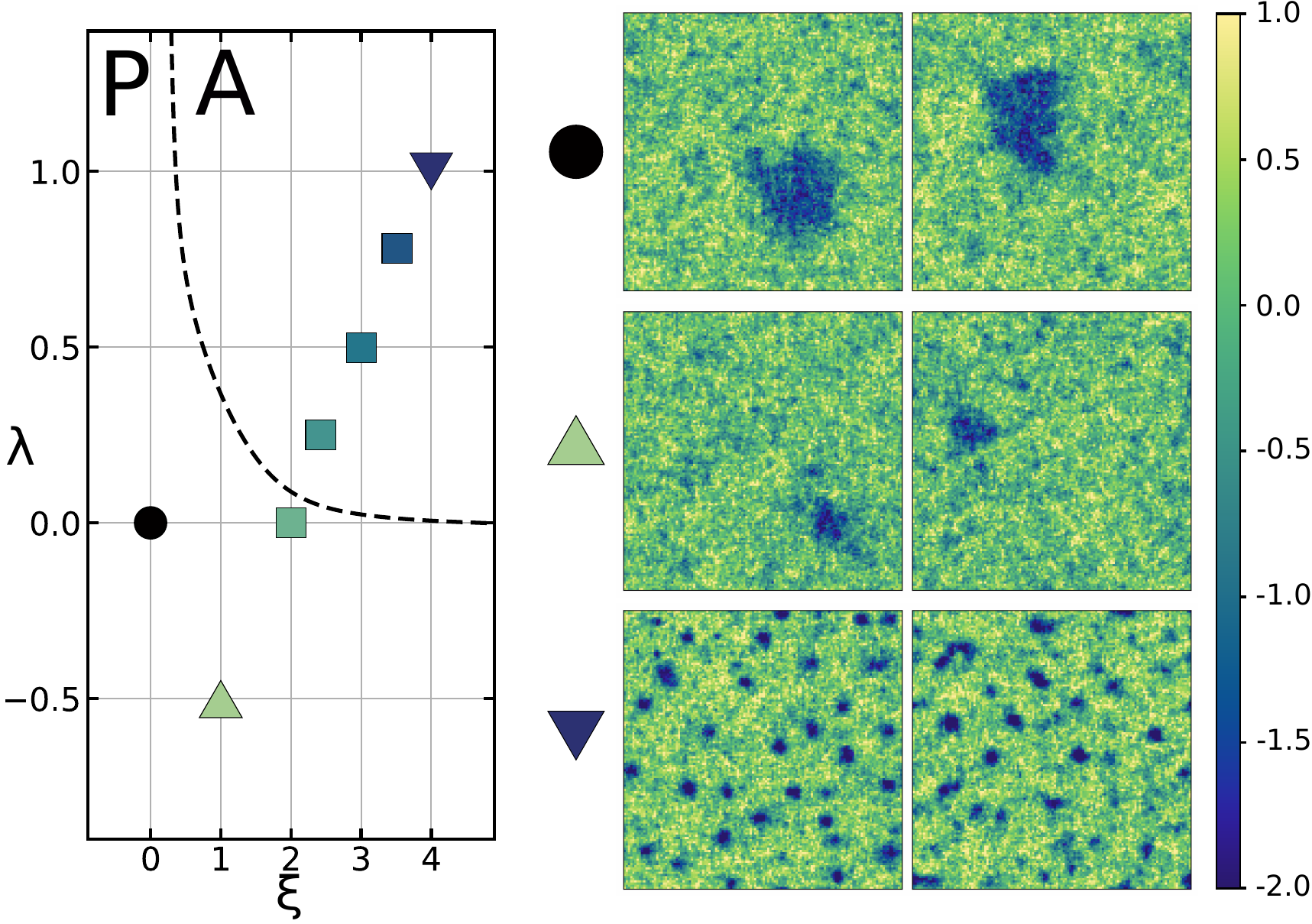}
    \caption{\textbf{Left:} Phase diagram of AMB+ in the noiseless regime $D=0$, adapted from \cite{tjhung_cluster_2018}. The P, for passive, phase is where the activity has no significant effect on the large scale phase separation, standard Ostwald ripening is observed. The A, for active, phase is where activity affects the large scale steady state phase separation and leads to the formation of microphases. Squares, triangle, and disk correspond to the set of parameters we considered to generate training dataset for the model. \textbf{Right:} For the three specific sets of parameters $(\lambda=0, \xi=0)$, $(\lambda=-0.5, \xi=1)$, $(\lambda=1, \xi=4)$, the field on the left is a representative example from the dataset, the one on the right is a sample generated from the learned WCRG model.}
    \label{fig:phase_diag}
\end{figure}

It has been shown that AMB+ can be obtained by coarse graining the microscopic dynamics of Active Ornstein-Uhlenbeck particles \cite{tjhung_cluster_2018}. Note that eq. (\ref{eq:amb_evo}) is a multi-dimensional Langevin equation in which the force, $- \nabla \cdot \mathbf{J} $, is not conservative. This is due to the parameters $\lambda$ and $\xi$, which control the activity of the model and break local detailed balance. The system does reach a steady state at long times, but it is a non-equilibrium one which does not verify the time-reversal symmetry (TRS). In this work all numerical results are obtained from a discretized square lattice version of eq. (\ref{eq:amb_evo}). The mean value of the field is conserved by the dynamics, we chose it equal to $0.6$ in order to focus on phase separation regimes. Additional details on the simulations are discussed in Methods \ref{app:amb}.

The main advantage of AMB+ in the context of this paper is that the amount of activity is tunable, via the two parameters $\lambda$ and $\xi$. Moreover, the model displays a phase transition which, depending on the values of $\lambda$ and $\xi$, can be equilibrium-like or show new features related to the violation of time-reversal symmetry \cite{caballero_bulk_2018,caballero_stealth_2020,fodor_how_2016}. 
Fig. \ref{fig:phase_diag} display the state points on which we have focused on in the $(\lambda,\xi)$ plane. In our study we have fixed $D=0.45$ which is below the transition temperature $D_c=0.54$ of the passive model, but still retain a substantial amount of thermal fluctuations.
We show in Fig. \ref{fig:phase_diag} the zero-noise $D=0$ separation line between the "effectively passive" and the "effectively active" phases, and snapshots of the corresponding phases (adapted from \cite{tjhung_cluster_2018}). 
For low level of activity, the large scale phase separation resembles what happens in equilibrium (top right panels). When the amount of activity increases the system display reversed Ostwald ripening and micro-phase separation (bottom right panels). For a more in depth discussion we refer the reader to \cite{tjhung_cluster_2018}.  

In the following we will apply the WCRG to the AMB+ in the different regions of the phase diagram. Our aim will be to identify the effective energy for each state point, and relate its main features to the non-equilibrium behavior of the system.

\section{Emergence of long-range interactions}

\begin{figure}[H]
    \centering
    \includegraphics[width=.9\linewidth]{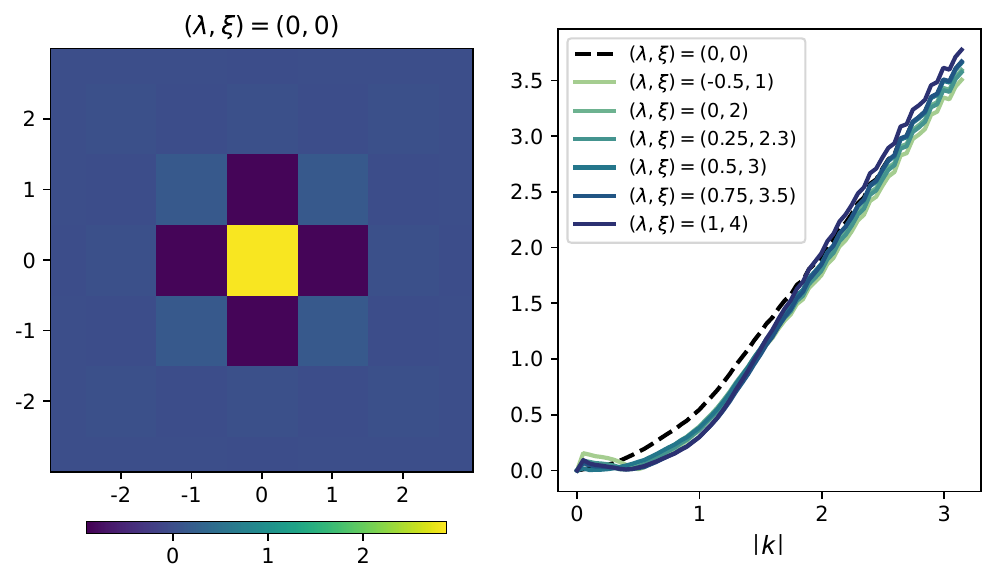}
    \caption{\textbf{Left:} Estimated Gaussian kernel $K_0$ for the passive setting. It is a discrete rotational invariant Laplacian. \textbf{Right:} Power spectrum of the kernel $K_0$, for different settings, as a function of wave number $k$. All kernels have almost the same spectrum.}
    \label{fig:Kernels}
\end{figure}

\begin{figure*}
    \centering
    \includegraphics[width=.99\linewidth]{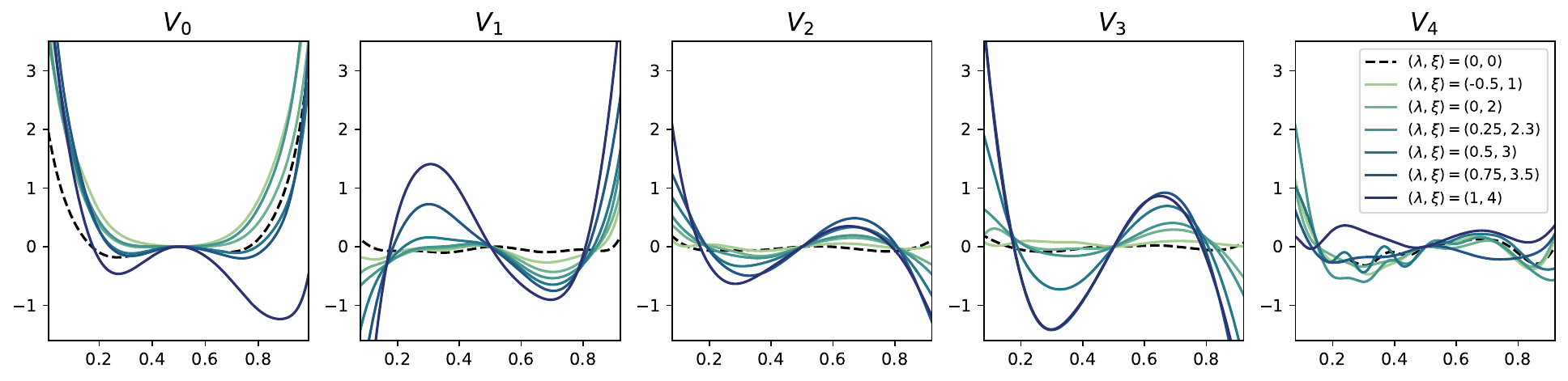}
    \caption{\textbf{(From left to right)} $\mbf{V_0}$\textbf{:} Finer scale scalar potential for different set of parameters. $\mbf{V_1}$, $\mbf{V_2}$, $\mbf{V_3}$ and $\mbf{V_4}$\textbf{:}  Multi-scale long range scalar potentials for different set of parameters, at increasing $j$. These potentials correspond to interactions on a length scale of $\ell_j=2^j$, so from left to right $\ell=1, \ell=2, \ell=4, \ell=8, \ell=16$. For the equilibrium system (dashed black line), only the microscopic $V_0$ potential is non zero as expected. The field $\varphi_j$ is rescaled to be in $(0,1)$ for all $j$.}
    \label{fig:multiscale_potentials}
\end{figure*}

We now present our first main result: the effective energy inferred from realisations of the system exhibits, when activity is strong, long range interactions which lead to micro-phase separation.

For each state point of Fig. 2, we have produced 5000 independent steady state configurations for a lattice of linear size $L=128$. Using these configurations as input, we have estimated the conditional probabilities, the quadratic kernels and the multi-scale potentials  introduced in equation \eqref{eq:ansatz}. For details on simulations and WCRG procedures see Methods.
Fig. \ref{fig:Kernels} presents results for the quadratic kernel $K_0$. On the left we show $K_0$ in real space for the passive case, which is at thermal equilibrium; it agrees with the discretized Laplacian $\Delta_{ij}$ we have used in the numerical simulations \footnote{The discretized Laplacian is $\Delta_{ij}=0.5(\delta_{i-1,j-1}+\delta_{i-1,j+1}+\delta_{i+1,j-1}+\delta_{i+1,j+1}) - 2(\delta_{i,j-1}+\delta_{i,j+1}+\delta_{i-1,j}+\delta_{i+1,j}) + 6 \delta_{i,j}$.}. On the right, we compare for all $(\lambda,\xi)$ of Fig. \ref{fig:phase_diag} the corresponding Fourier power spectra. They are numerically very close to the one corresponding to the passive case, thus showing that $K_0$ is a discretized Laplacian, or very close to it, for every  state point. The increase of activity has therefore no effect on this part of the effective energy. 

The situation is very different for the local potentials $V_j$. Each $V_j$ corresponds to a coarse-grained scale $\ell_j=2^{7-j}a$, where $a$ is the lattice spacing which henceforth we fix to one by adjusting the unit of length. As explained in Sec. \ref{subsec:WCRG}, each multi-scale potential encodes information about interactions on the scale $\ell_j$. For instance, $V_3$ corresponds to interactions  on a range corresponding to $\ell_3=8$ sites. Fig. \ref{fig:multiscale_potentials} shows $V_j$ for all different state points.

In the equilibrium case (black dashed line), we clearly see that the only significantly non-zero term is $V_0$ (the small non-zero values of $V_{j>0}$ are due to estimation errors). This is exactly what is expected:  the invariant probability distribution in equilibrium is $\exp(-\mathcal{F}/D)/Z_D$, and hence the effective potential $V_0$ coincides with the  one present in the dynamical equations through $\mathcal{F}$, whereas all other potentials are zero, as reported in the Supplementary Information. In equilibrium, short-range interactions in the dynamical equations imply short-range interactions in the energy function.  
For the cases with non-zero activity, the situation is different. Two main behaviors arise. For a relatively low level of activity (greener hues curves), which corresponds to systems behaving macroscopically similarly to the equilibrium case, as shown\footnote{Note the phase diagram showed here is the mean field one, so the actual transition line between the effectively passive and effectively active phase would be shifted.} in Fig.\ref{fig:phase_diag}, all the potentials, from $V_1$ and up, stay small, close to the zero baseline of the equilibrium case. Instead, for high activity (bluer hues curves), the potentials develop significant contributions up\footnote{$V_4$ and higher potentials are essentially zero except for atypical values of the field (for which the estimation is poor and the error large). Finite size effects are also likely to play a role. Their study would require to study and compare larger and smaller sizes, a problem which we leave for further works.} to $V_3$. These potentials are crucial to stabilize micro-phase separation. In fact, an energy function with a Laplacian $K_0$ and potentials as the $V_0$  and $V_1$ in Fig. \ref{fig:multiscale_potentials}, which locally favor one value of the field (the positive one), would lead to  macroscopic phase separation. The potentials $V_2$ and $V_3$ counteract the effect of $V_0$  and $V_1$ by favoring the other (negative) value of the field at a coarser level. The net result is a blocking of macroscopic phase separation, which is instead replaced by the formation of bubbles with length-scales up to $\ell_3$, as shown in Fig. \ref{fig:phase_diag}.
We find therefore two main results. 
First, a stabilization mechanism of  micro-phase separation which is different from the one found in many equilibrium theories \cite{leibler_theory_1980}. In the latter, the minimum of $K_0$ at a non-zero wave-vector leads through the quadratic term in the energy to micro-phase separation. In our case, instead, the form of $K_0$ is standard (it has a minimum at zero wave-vector), whereas they are the long-range coarse-grained potentials (not containing any quadratic contribution by construction), which play the crucial role.   
Second, once TRS is broken, even though forces are local, the interactions in the effective energy describing the non-equilibrium steady state become long-range for strong activity. Arguably, this is a key ingredient that allow the system to display phenomena which would be otherwise unexpected in equilibrium. 

We have checked the quality of the model estimated by WCRG, and its correctness testing that in the equilibrium passive case the energy function is directly linked to the forces in the dynamical equations. In the active case we have sampled scale by scale using the estimated conditional probabilities (running a Metropolis Adjusted Langevin diffusion \cite{besag_comments_1994}), and checked the quality of the image generated. Figure \ref{fig:phase_diag} shows side by side a true example of AMB+ extracted form the training dataset (on the left of the right panel) and a sample generated from the learned model for increasing level of activity. Clearly, in every case the generated samples are qualitatively very close to the dataset used to train on. This "visual correctness" is actually a difficult test to pass but it’s  important to also show quantitative comparisons. In the Supplementary Information (available online) we present more examples of samples from the model and also compare the Fourier spectrum and the probability distribution of the training and sampled datasets. All these show very good agreement in every situation tested.

\section{Out of equilibrium probes, length-scales, and patterns}\label{subsec:FDT}
We now present the second main result of this work: long-range interactions are directly associated to the out of equilibrium nature of the steady state. To this end, we study two ways to probe in details non equilibrium behavior. The first is the local entropy production rate. The knowledge of the effective energy enabled by the WRCG approach, allows to directly compute this quantity. The second is the violation of the Fluctuation Dissipation Theorem (FDT) as a function of the wave number.

\begin{figure*}
    \centering
    \includegraphics[width=.8\linewidth]{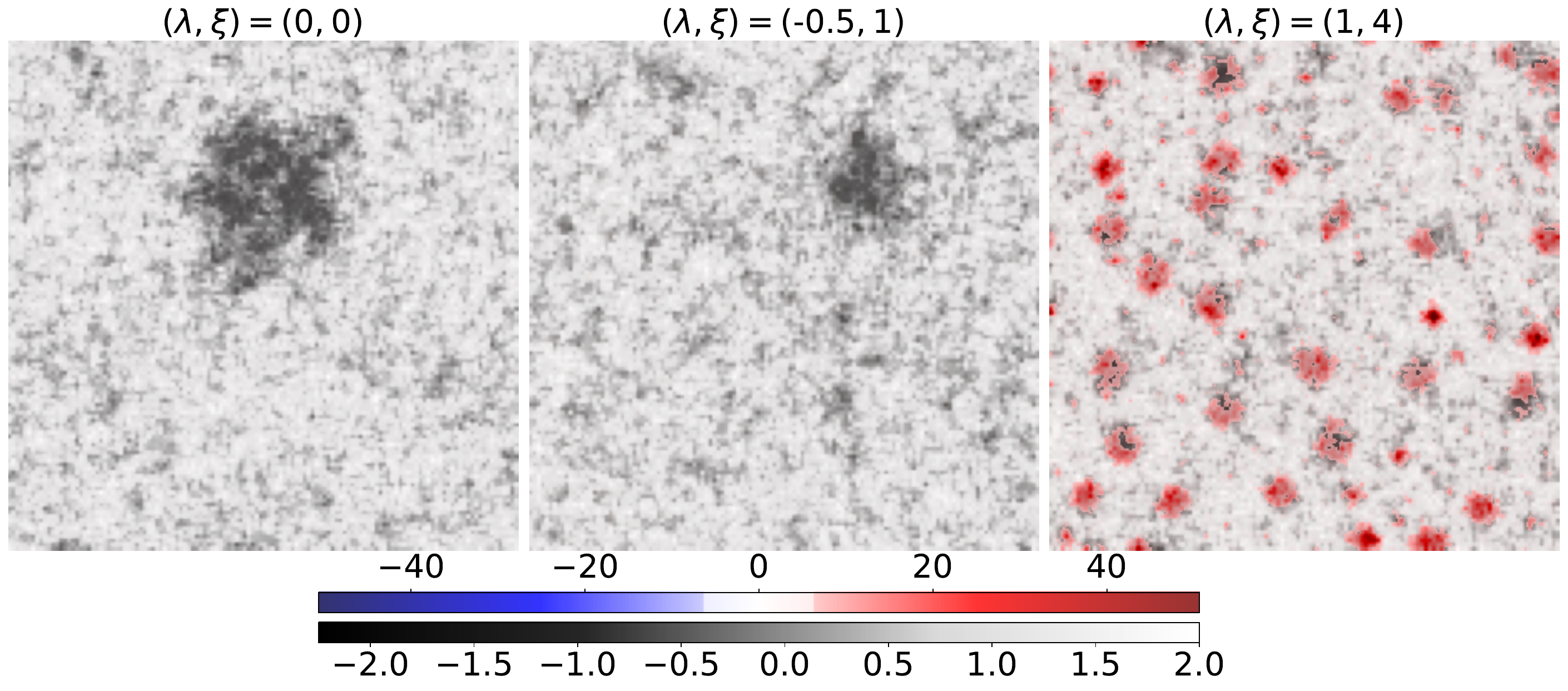}
    \caption{ Entropy production rate for \textbf{top left} equilibrium $(\lambda,\xi) = (0,0)$, \textbf{top right} below phase transition $(\lambda,\xi) = (-0.5,1)$ and \textbf{bottom left}  above phase transition $(\lambda,\xi) = (1,4)$ settings. The gray scale images encode the field value at each point. The superimposed color map shows the local entropy production rate, red for positive values, blue for negative ones and transparent close to zero. As expected the entropy rate is nearly everywhere positive or vanishing. The rate is significant only for a high degree of activity and stays minimal in the effectively passive regime. The entropy is produced mostly at the boundaries of the bubbles. This rate is obtained by computing the instantaneous production rate $\sigma$ and averaging it over a short time, short enough to not modify significantly the spatial structures.}
    \label{fig:entropy_prod}
\end{figure*}

Obtaining the patterns of entropy production \cite{seifert_stochastic_2012,cates_stochastic_2022} is in general a difficult task because it requires to know the gradient of the effective energy, a quantity which is unknown and difficult to obtain for non-equilibrium steady state. WCRG, by providing an estimation of $E_\text{eff}[\phi]$, offers a very direct way to probe the local entropy production field. In fact, once  the \textit{score} $\nabla_\phi \log p[\phi] = -\nabla_\phi E_\text{eff}[\phi]$ is known, using the force field $F= -\nabla_x \mathbf{J} $ and  the mobility matrix $\du{M}$ (coming from the discretization of the spatial differential operators), the local entropy production can be expressed as \cite{seifert_stochastic_2012, boffi_deep_2024}
\begin{equation}
    \sigma(\mathbf{x},t) = \du{M}(F + D\nabla E_\text{eff}) \cdot (F + D \nabla E_\text{eff}) \mdot
\end{equation}
More details on the derivation of this expression and important remarks on the computation of the different quantities can be found in Methods \ref{app:entropy_prod}. In Fig. \ref{fig:entropy_prod} we show $\sigma(\mathbf{x},t)$ (in red) superimposed on the field configuration (in grayscale) for realizations associated to three different state points. The left panel corresponds to the equilibrium case, the central one to an active system in a regime in which no long-range interactions are present in the effective energy. In both cases, the entropy production is absent or minimal. Note that in the case $(\lambda, \xi)=(-0.5,1)$ even though the microscopic dynamics contains terms breaking micro-reversibility, the system effectively behaves like if it was in equilibrium. 
The right panel corresponds to strong activity. In this case, entropy is instead produced locally in correspondence of the bubbles generating the micro-phase separation, thus on the scale of the effective long range interactions. It is mostly localized on the boundaries between the bubbles and the bulk\footnote{Note that the temperature we are investigating are not as low as in other studies, therefore the localization at the boundary is less sharp.}, as found by several previous studies \cite{nardini_entropy_2017,ro_play_2022}.

\begin{figure}[H]
    \centering
    \includegraphics[width=0.98\linewidth]{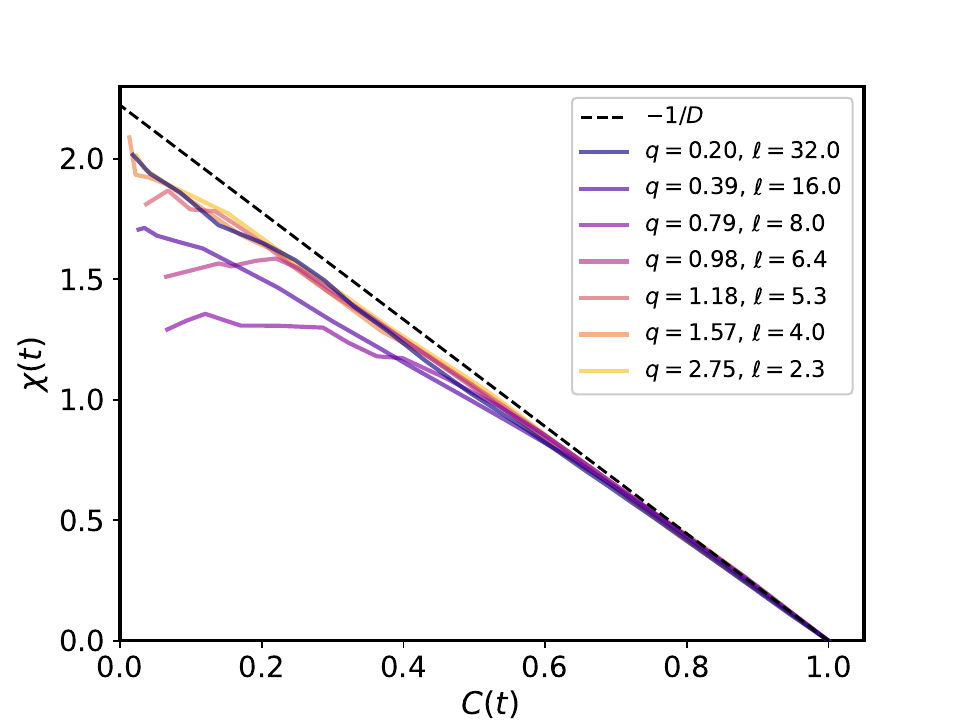}
    \caption{Parametric representation of the Fluctuation Dissipation Theorem at different wave numbers $q$ for $(\lambda, \xi)=(1,4)$. The typical length scale corresponding to wave number $q$ is denoted $\ell=2\pi/q$. The dashed line corresponds to the equilibrium FDT. The deviation is the strongest at intermediate $q$ values, which correspond to the size at which the effective potentials display long-range interactions and to the size of the bubbles in the micro-separated phase.}
    \label{fig:FDT}
\end{figure}

Analyzing the violation of the FDT has been a protocol used in several different physical situations to characterize out of equilibrium systems. In fact, FDT is directly related to the time-reversal symmetry of the dynamics. The way in which FDT is violated informs on the out of equilibrium nature of the system, as originally proven for spin-glasses \cite{cugliandolo_energy_1997}, and recently for active systems \cite{maggi_critical_2022}. One form of FDT is \cite{loi_effective_2008,tauber_critical_2014}
\begin{equation}
    \chi(q,t) = \frac{1}{D}\left[ C(q,0) - C(q,t) \right] ,
\end{equation}
where $C(q,t)$ and $\chi(q,t)$ are respectively the autocorrelation function and the integrated response function of the density field, at wave-vector $q$ and lag time $t$; $D$ is the equilibrium temperature. Following \cite{cugliandolo_energy_1997,maggi_critical_2022}, we study in Fig. \ref{fig:FDT} the parametric plot of $\chi(q,t)$ as a function of $C(q,t)$ for various values of $q$ in the case of strong activity. In equilibrium one should find a straight line with slope $-1/D$.

This is actually close to what we find for high and low spatial wave-vectors, i.e. small and large length-scales. Instead, when considering length-scales and state points at which the effective potentials display long-range interactions, we find that FDT is strongly violated. Concomitantly, the behavior of $C(q,t)$ is also peculiar, displaying a very slow relaxation for those modes $q$'s (see Fig. \ref{fig:AMB_correlation} in Methods). For state points corresponding to weak activity (effectively passive regime of Fig. \ref{fig:phase_diag}), we don't find any substantial violation of FDT for all wave-vectors (see SI). 

In summary, local entropy production patterns and FDT violation both highlight the connection between the effective out of equilibrium behavior and the emergence of long-range interactions in $E_\text{eff}$. For state points in which the system displays standard macroscopic phase separation, both out of equilibrium markers are absent or minimal and concomitantly the long-range interactions are not present. In this regime, the system is in a steady state which appears to be effectively in equilibrium despite the presence of terms breaking micro-reversibility in the equation of motion. A similar behavior is discussed in the case of active mixtures in \cite{dinelli_non-reciprocity_2023}. 
Instead, in the part of the phase diagram in which the systems displays micro-phase separation, entropy is produced and FDT is violated. Both probes indicate that the scale over which out of equilibrium effects are maximal correspond to the long range over which interactions identified in $E_\text{eff}$ act, and to the size of bubbles characterizing the micro-separated phase.

\section{Discussion}
The interpretable model of the effective energy obtained by the WCRG allows to connect the emergence of medium-long range interactions to characteristic out of equilibrium behaviors of a general model of scalar active matter, the AMB+.
Moreover, the form of the local potentials across scales offers a new perspective on the micro-phase separation out of equilibrium shown by active systems. 
Another application of WCRG which was not studied in this work but that would be worth exploring is a data-driven renormalization group theory for non-equilibrium steady states. In fact, by following the flow of  
$E_\text{eff}$ across scales one can characterize critical points and the associated relevant operators. It was shown in \cite{marchand_multiscale_2023} that for the standard equilibrium $\varphi^4$ field theory WCRG is indeed able to describe the RG critical fixed point. 

The key ingredient of WCRG is the ansatz for the form of the scale-dependent effective energy. The analysis of more complex active systems and of multi-scale out of equilibrium states, such as turbulence, will certainly require more general ansatzes. The proposal in \cite{lempereur_hierarchic_2024} could provide a robust low dimensional description sufficient to describe two-dimensional turbulence.
Modern machine learning generative methods applied to non-equilibrium physical systems, see e.g. \cite{li_stochastic_2024}, are very powerful but lacks interpretability, which is a key challenge to gain physical insights. The method presented in this paper is a step towards addressing this problem.

\appendix
\counterwithin{figure}{section}

\section{Methods}
\subsection{Wavelet bases}
\paragraph{Orthogonal Wavelet Filters}
We use an orthogonal decomposition that separates high from low frequencies, using two conjugate low and high pass discrete filters $(g,\overbar g)$.  We define the orthogonal operator $(G,\overbar G)$, with a convolution and a subsampling 
\begin{equation*}
\label{filtersGbarg}
    G \varphi (n) = \varphi * g(2n)~~\mbox{and} ~~\overbar G \varphi(n) = \varphi * \overbar g(2n) \mdot
\end{equation*}

In dimension $d$, conjugate mirror filters are computed as separable products of one-dimensional conjugate mirror filters $(g,\overbar g)$ \cite{mallat_wavelet_2009}. For a signal $\varphi$ of size $L^d$, $G$ computes a low frequency map of size $(L/2)^d$ whereas $\overbar G$ computes $2^d-1$ high frequency maps of size $(L/2)^d$. For images, $d=2$, $\overbar G$ computes vertical, horizontal and diagonal details.

\paragraph{Fast Wavelet transform}
We decompose the field $\varphi_0$ into its wavelet decomposition $\big(\overbar \varphi_j,\varphi_J\big)_{1 \leq j\leq J}$ by iteratively applying orthogonal operators $(G,\overbar G)$. The field $\varphi_{j-1}$, with length scale $2^{j-1}$, is decomposed into a coarser approximation $\varphi_j$, and $2^d-1$ wavelet coefficient $\overbar{\varphi}_j$ such that
\begin{equation*}
    \varphi_j = G\varphi_{j-1} ~~\rm{and}~~ \overbar\varphi_{j} = \overbar G\varphi_{j-1} \mdot
\end{equation*}
The finer scale field $\varphi_{j-1}$ can be recovered from $(\varphi_j,\overbar\varphi_j)$ with the transposed operators $(G^{\rm T},\overbar{G}^{\rm T})$ with
\begin{equation*}
    \varphi_{j-1} = G^{\rm T}\varphi_{j} + \overbar G^{\rm T}\varphi_{j} \mdot
\end{equation*}

\paragraph{Asymptotic wavelet bases}
When $j$ goes to $\infty$, for appropriate filters $\overbar g$ and low-pass filters $g$, one can prove \cite{daubechies_ten_1992} that the iterated wavelet filters $G_j = G^j$ and $\overbar G_j = G^{j-1}\overbar G$, such that $\varphi_j = G_j\varphi_{0}$ and $\overbar\varphi_j = \overbar G_j\varphi_{0}$, converges to $\phi(x)$ and wavelets $\psi_k (x)$, up to a dilation by $2^j$. These limit functions are square integrable. One can prove \cite{mallat_wavelet_2009} that $\{ 2^{-j/2} \psi_k(2^{-j} x -n) \}_{n \in \mathbb{Z^d} , j \in \mathbb{Z},1\leq k\leq d}$ is an orthonormal basis of $\mathcal{L}^2( \mathbb{R^d})$. Wavelet fields $\overbar\varphi_j$ can be rewritten as decomposition coefficients in this wavelet orthonormal basis. 

\paragraph{Choice of Wavelet basis}
The specific choice of wavelet $\psi$ made for the WCRG is important to optimize the performances of the method. Indeed with the development of wavelet theory, different kind of basis have been created to represent functions \cite{mallat_wavelet_2009}. One of the most important family of wavelets was invented by Daubechies \cite{daubechies_ten_1992} and bares her name. They have the property of having compact support and lead to orthogonal representation of $L^{2}(\mathbb{R})$ functions. One important feature is that they can be picked to have $p$ vanishing moments, for any $p \in \mathbb{N}$, which means the wavelet is orthogonal to any polynomial of degree $p-1$. There is a trade-off between the size of the support and the number of vanishing moments. Simplifying a bit, when trying to represent a field which is quite smooth, it is useful to increase the number of vanishing moments, because it will lead to a sparser wavelet representation \cite{mallat_wavelet_2009}. In the case of AMB+ presented in subsection \ref{subsec:AMB} inspection of the dynamical equations \eqref{eq:amb_evo} shows terms up to order $\nabla^4$ of the field. These high order gradient terms put strong constraints on the smoothness of the field and require the use of a wavelet with enough vanishing moment. We tested running WCRG using both Daubechies 4 symmlets (more symmetric version of standard Daubechies) and Daubechies 1 (also know as Haar) wavelets, the order indicating the number of vanishing moments. Using Daubechies 4 gave much better inference performances for the model in the case of AMB+. But the situation was reverse when looking at the simpler $\phi^4$ model which has lower order gradient terms, so not as much smoothness imposed by the dynamics. Furthermore, using a wavelet with too many vanishing moments leads to conditional energies much more complicated to model, with non-trivial interactions across scales, and thus approximation error \cite{lempereur_hierarchic_2024}.

\subsection{Score Matching, Free energy estimation and MALA algorithms for exponential families}
\paragraph{Exponential Families}
An exponential model $p_\theta$ of $p$ is defined from exponential models $p_{\theta_J}$ and $\overbar p_{\overbar \theta_j}$ of $p_J$ and $\overbar p_j$, from equation \eqref{eq:condiproba}
\begin{equation*}
\label{final-model}
    p_{\theta} (\varphi) = \alpha_0\, p_{\theta_J} (\varphi_J) \prod_{j=1}^{J} \overbar p_{\overbar \theta_{j}}(\overbar \varphi_{j}|\varphi_j) \mcom
\end{equation*}
where $\alpha_0$ is a normalization factor. 
An exponential model of $p_J$, with ansatz $\Phi_J$, is defined by
\begin{equation}
\label{internasf2}
    p_{\theta_J}(\varphi_J)  = {\cal Z}_J^{-1} e^{-\theta_J^{\rm{T}} \Phi_J(\varphi_J)} \mdot
\end{equation}
For any $j \leq J$, an exponential model of $\overbar p_j$, with ansatz $\Psi_j$, is defined by 
\begin{equation*}
\label{internasf}
    \overbar p_{\overbar \theta_j} (\overbar \varphi_j | \varphi_j) = e^{F_j (\varphi_j) - \overbar \theta_{j}^{\rm{T}} \Psi_j (\varphi_{j-1})} \mcom
\end{equation*}
where $F_j$ is a free energy which normalizes the conditional probability
\begin{equation*}
\label{normalisat-eq}
    \int \overbar p_{\overbar \theta_j} (\overbar \varphi_j | \varphi_j)\,  d \overbar \varphi_j = 
    e^{F_j (\varphi_j)} \int e^{- \overbar \theta_{j}^{\rm{T}} \Psi_j  (\varphi_{j-1})}\, d \overbar \varphi_j  = 1 \mdot
\end{equation*}
Each free energy $F_j$ is specified by $\overbar \theta_j$, but it does not need to be computed to estimate $\overbar \theta_j$ or sample $\overbar p_{\overbar \theta_j}$. $F_j$ is also approximated with as ansatz $\Phi_j$, such that $F_j\approx \alpha_j^{\rm T}\Phi_j$.

The model $p_\theta = {\cal Z}_\theta^{-1} e^{-E_\theta}$ has a Gibbs energy
\begin{equation}
\label{modelwithfree}
    E_{\theta} = \theta_J^{\rm{T}} \Phi_J + \sum_{j=1}^{J} \big(\overbar \theta_{j}^{\rm{T}} \Psi_j - \alpha_j^{\rm T}\Phi_j \big) \mdot
\end{equation}

\paragraph{Scalar potential ansatz}
At the coarsest scale $\Phi_j$ includes a two-point interaction matrix and a parametric scalar potential,
\begin{equation*}
\label{inter-param}
    \theta_J ^{\rm{T}} \Phi_J(\varphi_J) =\frac 1 2 \varphi_J^{\rm{T}}  K_J \varphi_J
    +  V_{\gamma_J} (\varphi_J) \mcom
\end{equation*}
The interaction Gibbs energy of $\overbar p_{\theta_j}( \overbar \varphi_j | \varphi_j)$ includes two-point interactions within the high frequencies $\overbar \varphi_j$, between high frequencies $\overbar \varphi_j$ and the lower frequencies $\varphi_j$, with convolution matrices $\overbar K_j$ and $\overbar K'_j$, plus a scalar potential
\begin{equation*}
    {\overbar \theta_{j}}^{\rm {T}}\Psi_j(\varphi_{j-1}) =
    \overbar \varphi_j^{\rm{T}}   \overbar K_{j}  \overbar \varphi_{j} +
    \overbar \varphi_j^{\rm{T}}   \overbar K'_{j}  \varphi_{j} +
    \overbar V_{\overbar \gamma_j} ( \varphi_{j-1}) \mdot
\end{equation*}
Finally, for the free energy ansatz, like for the coarsest scale
\begin{equation*}
    {\alpha_{j}}^{\rm {T}}\Phi_j (\varphi_{j-1}) =
    \varphi_j^{\rm{T}}   K_{j} \varphi_{j} +
    V_{\tilde\gamma_j} ( \varphi_{j}) \mdot
\end{equation*}
Under such parametrization, one can derive the scalar potential energy ansatz in equation \eqref{eq:ansatz} from equation \eqref{modelwithfree}.
    
Scalar potentials are averaged over sites, $V_\gamma (\varphi)= \sum_i v_{\gamma}(\varphi[i])$, with scalar potential $v_{\gamma}(t) = \sum_k \gamma_k\, \rho_k (t)$ decomposed over a finite approximation family $\{ \rho_k (t) \}_k$ with coefficients $\gamma = (\gamma_k )_k$.  We  use translated sigmoids: $\rho_k (t) = 1/(1+e^{(t-t_k)/\sigma_k)})$. In numerical applications, there are $25$ evenly spaced translations $t_k$, on the support of the distribution of each $\varphi_j(n)$, and $\sigma_k=\frac{3}{2}(t_{k+1}-t_k)$.
    
\paragraph{Score Matching}
Score matching provides a computationally scalable alternative to likelihood optimization of a model, which is valid for distributions with a bounded log-Sobolev constant, such as log-concave distributions. For conditional probabilities of scalar field theory, this hypothesis has been experimentally shown to hold \cite{marchand_multiscale_2023,guth_conditionally_2023,lempereur_hierarchic_2024}. Parameters $\overbar\theta_j$ are estimated by minimizing a relative Fisher information
\begin{small}
\begin{equation*}
    \ell(\overbar \theta_{j}) =  \mathbb{E}_{p_{j-1}}
    \Big(\|\nabla_{\overbar \varphi_{j}} \log \overbar p_{j} ( \overbar \varphi_{j}|\varphi_{j}) - \nabla_{\overbar \varphi_{j}} \log \overbar p_{\overbar \theta_{j}}
    ( \overbar \varphi_{j} | \varphi_{j})\|^2 \Big).
\end{equation*}
\end{small}
Following a derivation from \cite{hyvarinen_estimation_2005}, for an exponential family,
\begin{small}
\begin{equation*}
\label{eq:score}
    \ell(\overbar \theta_{j}) = \mathbb{E}_{p_{j-1}}
    \Big(\frac 1 2 \|  \overbar \theta_{j}^{\text T} \nabla_{\overbar \varphi_{j}} \Psi_j(\varphi_{j-1}) \|^2 - 
    \overbar \theta_{j}^{\text T}\Delta_{\overbar \varphi_{j}}  \Psi_j (\varphi_{j-1}) \Big),
\end{equation*}
\end{small}
from which we can derive the closed form
\begin{align*}
\label{eq:exact_theta_bar}
    \overbar\theta_j &= \overbar M_j^{-1}\mathbb{E}_{p_{j-1}}\Big(\Delta_{\overbar\varphi_j} \Psi_j(\varphi_{j-1})\Big) \mcom \\
    ~~ \text{with}~~ \overbar M_j &= \mathbb{E}_{p_{j-1}}\Big({\nabla_{\overbar\varphi_j} \Psi_j(\varphi_{j-1})\nabla_{\overbar\varphi_j }\Psi_j(\varphi_{j-1})^{\text T}}\Big) \mdot
\end{align*}
$M_j$ is an ill-conditioned quadratic matrix, which is regularized by adding $\epsilon\text{Id}$. $\theta_J$, at the coarsest scale, is inferred similarly. Due to the finite size of the datasets, empirical estimations replace the expectancies.

\paragraph{Free energy estimation}
By taking a derivative according to $\varphi_j$ in equation \eqref{internasf2}, the free energy and its parameterized approximation are shown \cite{lempereur_hierarchic_2024} to minimize the  quadratic loss function 
\begin{small}
\begin{equation*}
\label{cost-free}
    \ell( \alpha_{j}) = \mathbb{E}_{\overbar p_{\overbar \theta_{j}}p_j}\Big(||
    \alpha_{j}^{\text T}    \nabla_{\varphi_j} \Phi_j (\varphi_j)  -  { \overbar \theta_{j}^{\text T} \nabla_{\varphi_j}\Psi_j (\varphi_{j-1})} ||^2\Big) \mdot
\end{equation*}
\end{small}
We as well derive a closed form for $\alpha_j$
\begin{align*}
\label{eq:exact_theta_bar}
    \alpha_j = \tilde M_j^{-1}&\mathbb{E}_{\overbar p_{\theta_j}p_{j}}\Big(\nabla_{\varphi_j} \Phi_j(\varphi_{j}) \nabla_{\varphi_j} \Psi_j(\varphi_{j-1})^{\rm T}\Big)\overbar\theta_j \mcom \\
    ~~ \text{with}~~ \tilde M_j &= \mathbb{E}_{\overbar p_{\theta_j}p_{j}}\Big(\nabla_{\varphi_j} \Phi_j(\varphi_{j})\nabla_{\varphi_j }\Phi_j(\varphi_{j})^{\text T}\Big) \mdot
\end{align*}
We also regularize $\tilde M_j$.

\paragraph{Sampling with MALA}
We produce a sample $\varphi$ of $p_{\theta}$ from coarse to fine
\begin{itemize}
\item[$\blacksquare$] Initialization: compute a sample $\varphi_J$ of $p_{\theta_J}$. 
\item[$\blacksquare$] For $j$ from $J$ to $1$, given
$\varphi_j$ compute a sample $\overbar \varphi_j$ of $\overbar p_{\overbar \theta_j}( \cdot | \varphi_j)$ and set $\varphi_{j-1} = G \varphi_j + \overbar G \overbar \varphi_j$.
\end{itemize}

The sample $\varphi = \varphi_0$ of $p_{\theta}$ is obtained by iteratively sampling random high frequencies conditionally to low frequencies. Both $p_{\theta_J}$ and $\overbar p_{\overbar \theta_j}( \cdot| \varphi_j)$ are sampled using the Metropolis Adjusted Langevin Algorithm (MALA) \cite{besag_comments_1994}, which does not depend upon the normalization free energy $F_j$.

\paragraph{Regression of linear part}
Due to the fact that the dynamical evolution we consider conserve the spatial mean of the field, their is a gauge ambiguity in the scalar potential $V_j$. At each scale they are defined up to a linear part $a_j\varphi_j + b_j$. To remove this ambiguity, at each scale, we are regressing then subtracting away any linear part in the potentials obtained after the training procedure. It is these potentials which are presented in the main text in fig. \ref{fig:multiscale_potentials}. 

The gauge ambiguity emerges from the fact that for a given sample of the field, the sum over all lattice sites is equal to the same quantity (due to the conservation of the mean imposed by the dynamics). This property is also true for the coarse grained fields $\varphi_j$. This means that a linear term in the effective energy would lead to a field independent term. Indeed a linear term at scale $j$ is of the form $a_j\varphi_j + b_j = a_j \sum_x \varphi_j(x) + b_j = a_j m_j + b_j$ where $m_j$ is the average value of the field at the scale $j$, which is a fixed quantity for the system we consider.

\subsection{Active B+  Model simulation}\label{app:amb}
To produce the different training datasets the AMB+ was numerically simulated for different sets of parameters. The equation of motions \eqref{eq:amb_evo} where discretized (with $\Delta t = 0.01$, $\Delta x = \Delta y = 1.0$), then integrated using a simple temporal Euler scheme, where the white noise was discretize as a normally distributed random vector. Following \cite{tjhung_cluster_2018}, the different spatial operators where represented as finite difference approximations of high enough degree to properly capture the correct behavior. For each set of parameters the system was always looked at in the non-equilibrium steady state regime, by discarding the first samples until an estimated burn-in time was reached. The burning time was estimated by tracking some usual observables (correlation functions, moments of the field) and waiting long enough so they converge to constant steady state values. Periodic boundary conditions have been used in both directions.

\subsection{Correlation of AMB+}
\begin{figure}[H]
    \centering
    \includegraphics[width=0.99\linewidth]{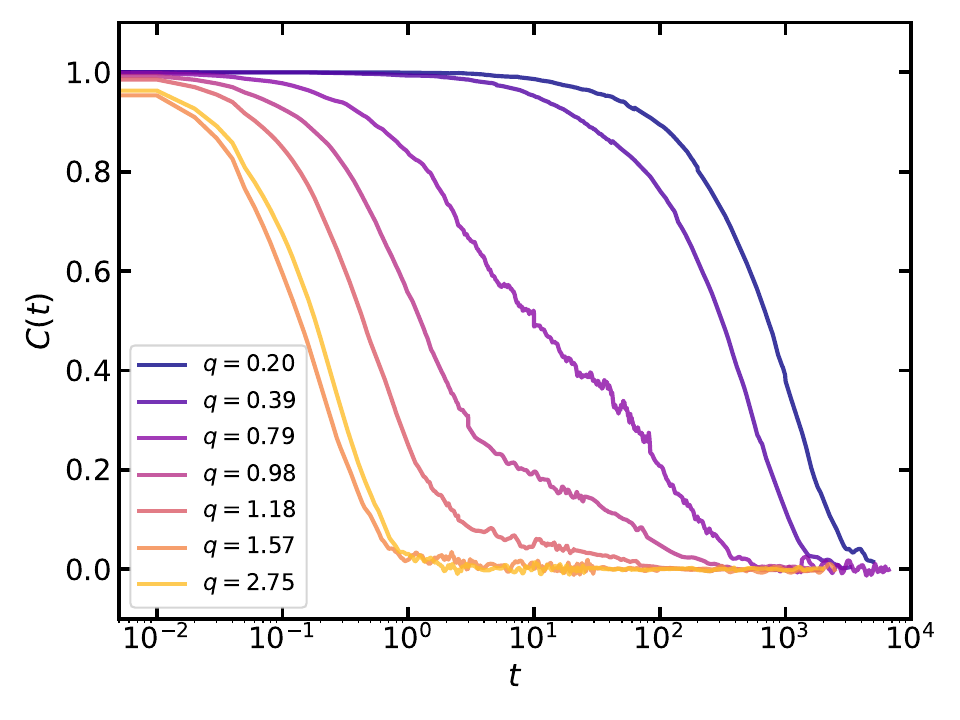}
    \caption{Spatial Fourier transform of the autocorrelation function of the AMB+ as a function of the time lag, for several wave numbers and for $(\lambda, \xi)=(1,4)$. Plotted on a semi-log scale. Low and high frequencies show a familiar exponential decay. However intermediate frequencies have a logarithmic relaxation.}
    \label{fig:AMB_correlation}
\end{figure}

We computed the autocorrelation of the AMB+ in the most active case considered in this study $(\lambda, \xi)=(1,4)$ for several wave numbers and over a wide range of time-lags, to see how the different modes decorrelate. Figure \ref{fig:AMB_correlation} illustrates once again the importance of the intermediate frequencies modes. Indeed contrary to the high and low frequencies modes which show a standard exponential decay of their autocorrelation, these intermediate mode have rather a logarithmic decay. This kind of behavior can be a marker of a wide range of out-of-equilibrium effects. In this case it is probably due to the strong long-range interactions acting on these modes.  

The autocorrelation for the effectively passive sector $(\lambda, \xi)=(-0.5,1)$, is available in the Supplementary Information and illustrates that in this case, all modes decay exponentially albeit with a large range of timescales.

\subsection{FDT, Malliavin weight for computing the response}
To show violation of the FDT, it is necessary to access the autocorrelation and the linear response function of the system. The autocorrelation is straightforward and can be estimated efficiently in the steady state by slicing the system in independant sequences and averaging over both these slices and several realisations of the system, to reduce somehow the overhead of letting the system equilibrate to steady state from the initial condition. The response is more difficult because naively it requires explicitly perturbing the system with an external field and then to compute an estimate of the derivative of the average value in the limit of a vanishingly small perturbation field. This is notoriously difficult. However alternative methods exist, which do not require to explicitly perturb the system. They rather introduce a supplementary variable, called the Malliavin Weight (due to its link with Malliavin calculus), which evolves alongside the system and allows to compute response to change of parameters \cite{warren_malliavin_2012, szamel_evaluating_2017}. This approach was already used to study active matter systems \cite{maggi_critical_2022}. However to exploit it in the context of the present paper, working with fields, a generalisation of the method is necessary. Extended details about the use of these approaches in the context of field theories will be made available in a future technical note, but the main result is the following. To compute the response function of AMB+, an auxiliary field is tracked during the simulation of the model. This Malliavin field $q_\alpha(x,t)$ evolves according to a Langevin equation 
\begin{equation*}
    \frac{\partial q_\alpha(x,t)}{\partial t} = \frac{1}{\sqrt{2D}} \int \diff{x'} \frac{\delta H_\alpha(x',t)}{\delta H_\alpha(x,t)} \nabla \cdot \Lambda(x',t) \mcom
\end{equation*}
where $H_\alpha$ is a external field added to the other forces in the dynamics for the field $\phi$ and $\Lambda$ is the same noise realisation as the one used to simulate the $\phi$ dynamics. Using the same realisation is crucial, because otherwise the Malliavin field would be uncorrelated to the field $\phi$. Equipped with $q_\alpha$ computing the response function simply amounts to taking the average of it times the field
\begin{equation*}
    \chi(x,t) = \avg{\phi(x,t) q_\alpha(x,t)} \mdot
\end{equation*}
It is straightforward to consider the Fourier transform of the Malliavin weight to compute the response as a function of the wave number.

\subsection{Entropy production}\label{app:entropy_prod}
To plot the entropy production in figure \ref{fig:entropy_prod}, we performed a short time average of successive time points to reduce the instantaneous noise in the entropy and have a more defined entropy production rate structure. These successive time points span an interval of $\Delta t = 0.1$. Comparing this time to the autocorrelation given in figure \ref{fig:AMB_correlation}, we this that this time interval is small enough to have nearly no relaxation of any of the modes. This means that the system stays basically the same, apart from very high frequency fluctuations coming from the stochastic noise, which is exactly what we wish to filter out.

\section*{Data availability}
The datasets used during this study are available from the corresponding authors upon request.

\section*{Code availability}
The code used to compute energies is derived from the code of \cite{lempereur_hierarchic_2024} and available in the repository \url{https://github.com/Elempereur/WCRG}. The code used to simulate AMB+ is adapted from \cite{tjhung_cluster_2018}. The code used for FDT and entropy production is available from the corresponding authors upon request.

\section*{Acknowledgments}
We warmly thank Misaki Ozawa for valuable discussions, and for participating to the initial stage of this work. We also thank Nicoletta Gnan and Claudio Maggi for discussions, and pointing out their studies of FDT violation for active matter systems, and Julien Tailleur for helpful inputs on interactions in active systems.

\section*{Author contributions}
A.B.: conceptualization, numerical experiment, writing, reviewing. 
E.L: conceptualization, numerical experiment, writing, reviewing. 
S.M.: conceptualization, writing, reviewing, supervision. 
G.B.: conceptualization, writing, reviewing, supervision. 

\end{multicols}
\printbibliography
\end{document}